\documentclass{article}

\usepackage{color, graphicx}

\usepackage{hyperref}
\usepackage{amsmath, amssymb, amsthm, amsfonts}
\usepackage[margin=1.0in]{geometry}

\newcommand{\Prob}{\mathrm{Prob}}

\newtheorem{thm}{Theorem}

\newtheorem{cor}[thm]{Corollary}

\theoremstyle{remark}
\newtheorem{rem}[thm]{Remark}

\title{On the combinatorics of last passage percolation in a quarter square and $\mathrm{GOE}^2$ fluctuations}

\author{
  Dan Betea~\thanks{Institute for Applied Mathematics, University of Bonn, \texttt{dan.betea@gmail.com}} 
}

\begin{document}

\maketitle

\begin{abstract}
    In this note we give a(nother) combinatorial proof of an old result of Baik--Rains: that for appropriately considered independent geometric weights, the generating series for last passage percolation polymers in a $2n \times n \times n$ quarter square (point-to-half-line-reflected geometry) splits as the product of two simpler generating series---that for last passage percolation polymers in a point-to-line geometry and that for last passage percolation in a point-to-point-reflected (half-space) geometry, the latter both in an $n \times n \times n$ triangle. Equivalently, the probability distribution of the first is a product of the latter two distributions. As a corollary, we immediately recover a discrete version of a result of Baik--Rains: that for iid geometric random variables---of parameter $q$ off-diagonal and parameter $\sqrt{q}$ on the diagonal---the last passage percolation time in said quarter square obeys Tracy--Widom $\mathrm{GOE}^2$ fluctuations in the large $n$ limit as both the point-to-line and the point-to-point-reflected geometries have known GOE fluctuations. 
\end{abstract}

\section{Introduction}

\paragraph{Background and motivation.} 
Discrete point-to-point last passage percolation with iid geometric weights of parameter $q$ was introduced by Johansson~\cite{joh2}. Using the Robinson--Schensted--Knuth~\cite{knu} correspondence, Johansson showed that the last passage time is a certain observable of a large class of determinantal measures called Schur measures~\cite{oko}. Using this together with orthogonal polynomial techniques, he then analyzed the last passage time asymptotically to find the celebrated KPZ $n^{1/3}$ fluctuation scaling regime and the Tracy--Widom~\cite{tw} GUE distribution as the limiting distribution for fluctuations. The result also holds in the $q \to 0$ limit (the Poisson case) and the $q \to 1$ limit (the case of iid exponential weights).

Other geometries have also been considered. Rains~\cite{rai}, Baik--Rains~\cite{br1, br2}, Ferrari~\cite{PF04}, Forrester--Rains~\cite{fr} and Bisi--Zygouras~\cite{bz1, bz2} have considered the similar problem in a point-to-line geometry and here the last passage time is an observable of a \emph{pfaffian} Schur measure (respectively \emph{symplectic measure} for Bisi--Zygouras). The last passage percolation time (at least in the Poisson and exponential limits) has been shown to have Tracy--Widom GOE fluctuations in the works of Baik--Rains~\cite{br2} (Poisson limit), Ferrari and Bisi--Zygouras~\cite{PF04, bz2} (exponential limit). Rains~\cite{rai}, Baik--Rains~\cite{br1, br2}, Forrester--Rains~\cite{fr}, Sasamoto--Imamura~\cite{si}, Baik--Barraquand--Corwin--Suidan \cite{BBCS17} and Betea--Bouttier--Nejjar--Vuleti\'c~\cite{bbnv} have further considered the point-to-point-reflected (sometimes dubbed half-space) geometry where again for geometric/Poisson/exponential weights, the fluctuations of the last passage time are Tracy--Widom GOE~\cite{TW96}\footnote{In the case, say, of independent geometric weights, we want the parameter for the weights on the diagonal be the square root of the off-diagonal parameter.}.

\begin{figure}[!ht] 
    \begin{center}
    \includegraphics[scale=1]{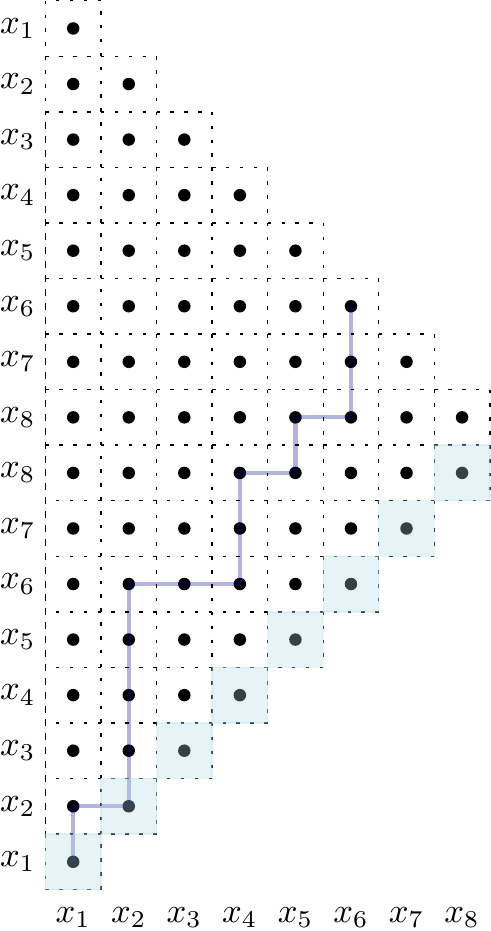} \\ \ \\ \ \\    
    \includegraphics[scale=1]{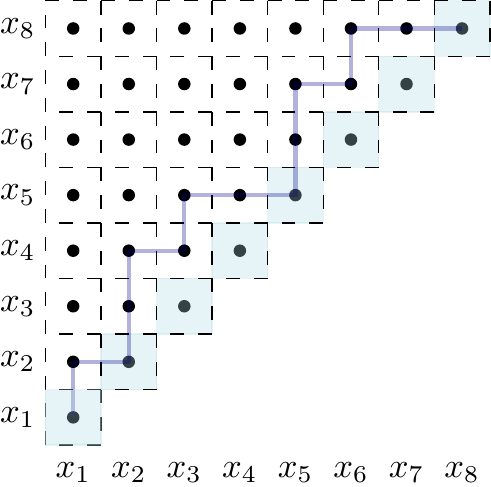} \qquad \qquad \includegraphics[scale=1]{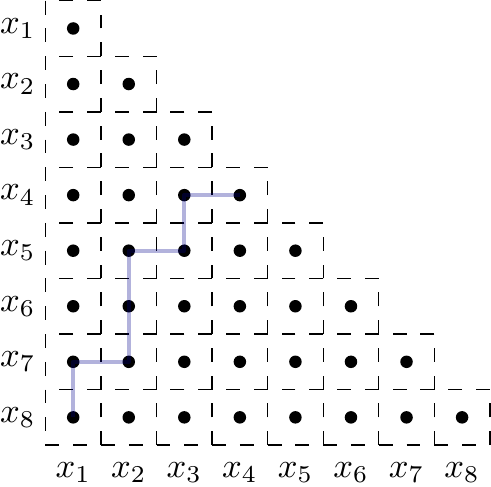}
\end{center}
    \caption{The three geometries and types of polymers studied in this paper (depicted for $n=8$): point-to-line-reflected (top), point-to-point-reflected (bottom left) and point-to-line (bottom right). The light blue squares act as reflecting boundaries. The bullets are independent non-negative integer valued random variables---with parameter $Geom(x_i x_j)$ away from the diagonal (blue squares) and parameter $Geom(x_i)$ on the diagonal (reflecting boundaries, only in the first two cases). Example polymers (up-right paths) are depicted in dark blue, the weights of which are the sums of the numbers (bullets) collected.}
    \label{fig:ex_polymers}
\end{figure}

In this note we consider last passage percolation in a quarter square which we call, inspired by the previous paragraph, \emph{point-to-half-line-reflected geometry}---see Figure~\ref{fig:ex_polymers} (top)---with independent geometric weights and show that the distribution for the last passage percolation time is a product of two distributions: that of the time in a point-to-line geometry and that in a point-to-point-reflected (half-space) geometry---both of which were discussed in the previous paragraph. This fact leads to $\mathrm{GEO}^2$ fluctuations in the appropriate $n^{1/3}$ fluctuation regime when all weights are iid geometric random variables with parameter $q$ off-diagonal and $\sqrt{q}$ on diagonal. This model was considered and different formulas for the distribution first obtained by Rains~\cite{rai}, Baik--Rains~\cite{br1} (for the Poisson limit $q \to 0$), Rains--Forrester~\cite{fr}, and Bisi--Zygouras~\cite{bz1} (for the exponential limit $q \to 1$). Asymptotics revealing $\mathrm{GOE}^2$ fluctuations were considered by Baik--Rains~\cite{br2} (for the Poisson limit, using Toeplitz+Hankel determinants and Riemann--Hilbert techniques) and Bisi~\cite{bis} (for the exponential limit, using classical steepest descent analysis). 

One of the upshots of the present note is that, like in the case of Baik--Rains for the Poisson limit~\cite{br1, br2}, no analysis is required to arrive at said Tracy--Widom $\mathrm{GOE}^2$ fluctuations. The mere fact that the distribution splits as a product of two known simpler distributions with known asymptotics is enough to pass to the limit in the original one. Another, more subtle, is that the result as proven combines bijective and representation theoretic techniques---notably Robinson--Schensted--Knuth correspondences and bounded Littlewood identities for various types of characters---which could hopefully be amenable to generalizations studying so-called ``non-free fermionic models''. 

\paragraph{The main result.} 
To state the main result, let us fix some notation and vocabulary. Since one of the main ingredients in the proof is a discrete version of a result of Bisi--Zygouras~\cite{bz1}, we will keep notation and terminology close to op. cit. whenever feasible. Throughout, $n$ and $u$ will denote positive integers, and we will assume---to simplify some formulas---that $u := 2v$ is even. Furthermore, we will use $x_1, \dots, x_n$ as variables throughout. 

Consider the following three discrete domains in the quarter plane $i,j \geq 0$ ($i$ the horizontal axis) built up from $1 \times 1$ unit squares: $D^{\mathrm{p2hlr}}_n$ is the $2n \times n \times n$ discrete quarter square (triangle) consisting of the $n^2+n$ unit squares on or above the diagonal $i=j$ and on or below the anti-diagonal $i = 2n-j+1$; $D^{\mathrm{p2pr}}_n$ is the $n \times n \times n$ triangle consisting of the $n(n+1)/2$ unit squares on or above the diagonal $i=j$ and below the horizontal line $j = n$; and $D^{\mathrm{p2l}}_n$ is the $n \times n \times n$ triangle consisting of the $n(n+1)/2$ unit squares on or below the anti-diagonal $i=n-j+1$. They are depicted in Figure~\ref{fig:ex_polymers} (top, bottom left and bottom right respectively). The meaning of the abbreviations stands for the type of geometry we will consider: p2hlr stands for \emph{point-to-half-line-reflected}, p2pr stands for \emph{point-to-point-reflected}, and p2l stands for \emph{point-to-line}.

Each unit square $(i, j)$\footnote{$(i, j)$ is the cartesian coordinate of the top right corner of the square.} in each of the three types of triangles has a non-negative integer $w_{i,j}$ sitting inside it, and to such a filling $W:= (w_{i,j})$ we associate a weight $\mathrm{wt}(W)$ as follows. The total weight of a triangle is a product over weights of all individual unit squares inside it. A unit square $(i, j)$ contributes a weight $(x_i x_j)^{w_{i,j}}$ if it off-diagonal and---in the case of triangles from $D^{\mathrm{p2pr}}_n$ and $D^{\mathrm{p2hlr}}_n$ only---a weight $x_i^{w_{i,j}}$ if the square is on-diagonal. The parameters are assigned to each row and column inside the domains thusly: $x_i$ corresponds to column $i$ (placed under the $i$-axes in Figure~\ref{fig:ex_polymers}); while $x_j$ corresponds to row $j$ in the point-to-point-reflected (p2pr) geometry, to row $n-j+1$ in the point-to-line geometry (p2l) and to rows $j$ and $2n-j+1$ in the point-to-half-line. We are interested in: the set of all fillings $W$ of domain $D^{\mathrm{p2hlr}}_n$ such that the longest up-right path---called a \emph{polymer}---from the $(1,1)$ square to any square on the line $i = 2n-j+1$ is $\leq u$; the set of all fillings $W$ of domain $D^{\mathrm{p2pr}}_n$ such that the longest up-right path/polymer from the $(1,1)$ square to the $(n, n)$ square is $u$; and the set of all fillings $W$ of domain $D^{\mathrm{p2l}}_n$ such that the longest polymer from the $(1,1)$ square to any square on the line $i=n-j+1$ is $\leq v = u/2$. Here the \emph{length} of a polymer is the sum of all $w_{i,j}$ crossed by the associated path and the longest such polymer is usually called the \emph{last passage percolation (LPP) time} for that geometry. We denote the latter by $L^{\mathrm{p2hlr}}_n$, $L^{\mathrm{p2pr}}_n$ and $L^{\mathrm{p2l}}_n$. We are thus interested in the sets $\mathcal{W}^{\mathrm{x}}_{n, \ell} = \{ W | L^{\mathrm{x}}_n \leq \ell \}$ where $\mathrm{x} \in \{ \mathrm{p2hlr}, \mathrm{p2pr}, \mathrm{p2l} \}$ and $\ell = u$ if $\mathrm{x} \in \{ \mathrm{p2hlr}, \mathrm{p2pr} \}$ and $\ell=v=u/2$ otherwise. See Figure~\ref{fig:ex_polymers} for the geometry, sample polymers and parametrization of rows and columns---note the weights $w_{i, j}$ are replaced by bullets.

The main result is the following equality of generating series. 
\begin{thm} \label{thm:main_thm}
    We have:
    \begin{equation}
        \sum_{W \in \mathcal{W}^{\mathrm{p2hlr}}_{n, u}} \mathrm{wt}(W) = \left( \sum_{W \in \mathcal{W}^{\mathrm{p2pr}}_{n, u}} \mathrm{wt}(W) \right) \cdot \left( \sum_{W \in \mathcal{W}^{\mathrm{p2l}}_{n, v}} \mathrm{wt}(W) \right).
    \end{equation}
\end{thm}

In other words and with parameters as stated above, the generating series for point-to-half-line-reflected polymers of length at most $u$ is a product of the two generating for point-to-point-reflected and point-to-line polymers of length at most $u$ and respectively $v=u/2$. 

The result will be proven using purely combinatorial techniques and formulas pieced together from the works of Okada~\cite{oka2}, Stembridge~\cite{ste} and Bisi--Zygouras~\cite{bz1}. 

\begin{rem} \label{rem:other_proof}
Theorem~\ref{thm:main_thm} is implicit already in the work of Baik--Rains~\cite[Corollary 4.3]{br1}: one can multiply equations (4.21) and (4.23) to obtain equation (4.25) in op. cit., which certainly implies the result. We believe the non-intersecting lattice paths proof of Forrester--Rains~\cite[Section 6]{fr} can also be modified to produce the same outcome. The idea in both references is that the left-hand side in Theorem~\ref{thm:main_thm} can be written as a sum over semi-standard \emph{domino} tableaux which are in bijection to pairs of semi-standard Young tableaux which then yield a product of two bounded Littlewood sums of Schur polynomials giving the product on the right-hand side. A similar idea is employed in this note except instead of passing through self-dual (in the Sch\"utzenberger sense) and domino tableaux, we go through symplectic and orthogonal tableaux and characters. We thus side-step the Sch\"utzenberger involution~\cite{ful}\footnote{Or at least sweep it under a moderately thick rug.}.
\end{rem}

Passing to probability, suppose $x_i = \sqrt{q}$ for all $i$ where $0 < q < 1$ is a fixed parameter and suppose $w_{i,j}$ are now independent geometric random variables of parameter $q$ off-diagonal and $\sqrt{q}$ on-diagonal for the p2hlr and p2pr geometries\footnote{$X$ is a geometric random variable of parameter $q$ on $\mathbb{N}$ if $\Prob(X=k) = (1-q) q^k$.}. Based on known $n^{1/3}$ asymptotics for the p2pr and p2l geometries---see Section~\ref{sec:main_proof} for details and references, we have the following corollary.

\begin{cor} We have:
    \label{cor:main_cor}
    \begin{equation}
        \lim_{n \to \infty} \Prob\left(\frac{L^{\mathrm{p2hlr}}_n - c_1 n}{c_2 n^{1/3}} \leq s \right) = F_1^2(s)
    \end{equation}
    where
    \begin{equation}
        \label{eq:constants}
        c_1 = \frac{2 \sqrt{q}}{1-\sqrt{q}}, \qquad c_2 = \frac{q^{1/6} (1+\sqrt{q})^{1/3}}{1-\sqrt{q}}
    \end{equation}
    and where $F_1(s)$ is the Tracy--Widom GOE distribution~\cite{TW96}.
\end{cor}

\begin{rem}
Corollary~\ref{cor:main_cor} appears as Theorem 4.2 equation (4.30) with ($w = 0, \beta=0$ case) in the extended abstract~\cite{br99}, with the proof left to the reader but following along similar lines as the proof of the Poisson limit $q \to 0$. This latter Poisson limit case of Corollary~\ref{cor:main_cor} was proven in Baik--Rains~\cite{br2} based on the combinatorics explained in Remark~\ref{rem:other_proof} and on Riemann--Hilbert techniques for the asymptotics of certain Toeplitz+Hankel determinants. The corollary has also been recently obtained by Bisi (personal communication and~\cite{bis2}) using classical (as opposed to Riemann--Hilbert) steepest descent analysis on the corresponding pfaffian distribution from~\cite{bz1}.
\end{rem}

\paragraph{Organization of the paper.} 
The paper is organized as follows. In Section~\ref{sec:proofs} we prove the main result. We first set up the machinery of Schur functions and symplectic/orthogonal characters along with the associated tableaux/Gelfand--Tsetlin patterns in Section~\ref{sec:prelims} and then prove the main result in Section~\ref{sec:main_proof}. We conclude in Section~\ref{sec:conclusion}. As the heavy lifting involves variants of the Robinson--Schensted--Knuth correspondence, we recall all the required material in Appendix~\ref{sec:appendix_rsk} using the well-established language of Fomin (corner) growth diagrams. Finally, to prevent the somewhat large figures from interrupting the flow of text, we have placed all figures at the end of the document.

\paragraph{Acknowledgements.} 
The author acknowledges illuminating conversations with Nikos Zygouras and Elia Bisi regarding~\cite{bz1}; equally illuminating conversations with Peter Nejjar leading to the correction of a glaring ``off-by-two'' error in a previous version of this manuscript and with Patrik Ferrari on the innards of LPP models; and finally Elia Bisi in particular for providing the author with a copy of his PhD thesis. 

\section{Proofs}
\label{sec:proofs}

\subsection{Some preliminaries}
\label{sec:prelims}

An (integer) \emph{partition} $\lambda$ is a non-increasing sequence of non-negative integers $\lambda_1 \geq \lambda_2 \geq \cdots$, only finitely many of them non-zero. The non-zero integers are called \emph{parts}. The \emph{length} of $\lambda$, denoted $\ell(\lambda)$, is the number of its (non-zero) parts, while its \emph{size}, denoted $|\lambda|$, is the sum of its parts $|\lambda|:=\sum_{i \geq 1} \lambda_i$. The \emph{empty partition}, denoted by $\emptyset$, is the one with no parts and has length 0. Partitions are usually written/thought of as \emph{Young diagrams}, collections of square boxes, left aligned, containing $\lambda_i$ squares in row $i$, counting from the top. For two partitions $\mu \subset \lambda$---meaning $\mu_i \leq \lambda_i \ \forall i$, we write $\mu \prec \lambda$ (respectively $\mu \prec' \lambda$) and say $\mu$ \emph{(upwards) interlaces with} $\lambda$ (respectively $\mu$ \emph{dual interlaces with} $\lambda$), if $\lambda_i \geq \mu_i \geq \lambda_{i+1}$ (respectively $\lambda_i - \mu_i \in \{0,1\}$) for all $i$. 

A \emph{Gelfand--Tsetlin pattern} of height $n$ is a triangular array $z = (z_{i,j})_{1 \leq j\leq i \leq n}$ with non-negative integer entries that satisfy the \emph{interlacing conditions} $z_{i+1,j+1} \leq z_{i,j} \leq z_{i+1,j}$ for all meaningful $i, j$. Its \emph{shape} is the bottom row partition $(z_{n,1}, \dots, z_{n,n})$. Its \emph{type} is the vector $\mathrm{type}(z)$ defined by $\mathrm{type}(z)_i = \sum_{j = 1}^i z_{i,j} - \sum_{j = 1}^{i-1} z_{i-1,j}$ for $1 \leq i \leq n$. A Gelfand--Tsetlin pattern $z$ of height $n$ and shape $\lambda$ can be equivalently viewed as an \emph{upwards interlacing sequence of $n+1$ partitions}
\begin{equation} 
    z \leftrightarrow \Lambda = \left(\ \emptyset = \lambda^{(0)} \prec \lambda^{(1)} \prec \dots \prec \lambda^{(n)} = \lambda \right)
\end{equation}
via $\lambda^{(i)}_j := z_{i,j}$. Note $\ell(\lambda^{(i)}) \leq i$. Finally by placing the symbol $i$ ($1 \leq i \leq n$) in the \emph{skew Young diagram} $\lambda^{(i)}/\lambda^{(i-1)}$---the boxes in $\lambda^{(i)}$ and not in $\lambda^{(i-1)}$, one obtains a third equivalent characterization: \emph{a semi-standard Young tableau (SSYT)} $T$---a filling of $\lambda$ with symbols from $1 < 2 < \dots < n$ strictly increasing down columns and weakly increasing down rows. Observe that in the equivalence $z \leftrightarrow \Lambda \leftrightarrow T$ we have $\mathrm{type}(z)_i = |\lambda^{(i)}| - |\lambda^{(i-1)}| = \text{number of $i$ in $T$}$.

Given an integer partition $\lambda$ with $\ell(\lambda) \leq n$, we denote by $GT_n(\lambda)$ the set of all Gelfand--Tsetlin patterns of height $n$ and shape $\lambda$.

A \emph{symplectic Gelfand--Tsetlin pattern} of height $2n$ is a ``half-triangular'' array $z=(z_{i,j})_{1\leq i\leq 2n, 1\leq j\leq \lceil i/2 \rceil}$ with non-negative integer entries satisfying the interlacing conditions $z_{i+1,j+1} \leq z_{i,j} \leq z_{i+1,j}$ for $1\leq i < 2n, 1\leq j\leq \lceil i/2 \rceil$ with the convention that $z_{i,j}:=0$ when $j>\lceil i/2 \rceil$ (so that all its entries are non-negative). Its \emph{shape} is the bottom row partition $(z_{2n,1},\dots,z_{2n,n})$, and its \emph{type} the vector $\mathrm{type}(z)$ with $\mathrm{type}(z)_i := \sum_{j = 1}^{\lceil i/2 \rceil} z_{i,j} - \sum_{j = 1}^{\lceil (i-1)/2 \rceil} z_{i-1,j}$ for $1 \leq i \leq 2n$.  A symplectic Gelfand--Tsetlin pattern $z$ of height $2n$ and shape $\lambda$ can be equivalently viewed as an \emph{upwards interlacing sequence of $2n+1$ partitions}
\begin{equation} 
    z \leftrightarrow \Lambda = \left(\ \emptyset = \lambda^{(0)} \prec \lambda^{(1)} \prec \dots \prec \lambda^{(2n)} = \lambda \right)
\end{equation}
via $\lambda^{(i)}_j := z_{i,j}$. Note $\ell(\lambda^{(i)}) \leq \lceil i/2 \rceil$. Finally by placing the symbol $i$ respectively $\overline{i}$ ($1 \leq i \leq n$) in the \emph{skew Young diagram} $\lambda^{(2i-1)}/\lambda^{(2i-2)}$ respectively $\lambda^{(2i)}/\lambda^{(2i-1)}$, one equivalently obtains a \emph{symplectic tableau (SpT)} $T$ of King~\cite{kin,kel}---a semi-standard Young tableau of shape $\lambda$ on $1 < \overline{1} < 2 < \overline{2} < \dots < n < \overline{n}$ having the \emph{symplectic property}: elements in row $i$ are $\geq i$. In the equivalence $z \leftrightarrow \Lambda \leftrightarrow T$ we have $\mathrm{type}(z)_{2i-1} = |\lambda^{(2i-1)}| - |\lambda^{(2i-2)}| = \text{number of $i$ in $T$}$ and $\mathrm{type}(z)_{2i} = |\lambda^{(2i)}| - |\lambda^{(2i-1)}| = \text{number of $\overline{i}$ in $T$}$. An example is given in Figure~\ref{fig:p2hlr} (bottom right).

A Sundaram~\cite{sun3} \emph{odd orthogonal tableau} (OOT) of shape $\lambda$ and height $n$ is a filling of $\lambda$ with the alphabet $1 < \overline{1} < \overline{2} < \dots < n < \overline{n} < \infty$ satisfying the same conditions as a symplectic tableau and the extra condition that no two $\infty$ symbols can appear in the same line of the Young diagram $\lambda$. As above, it can be put into a correspondence with a sequence of $2n+2$ partitions, the first $2n+1$ interlacing (and the last two \emph{dual interlacing}) or a \emph{odd orthogonal Gelfand--Tsetlin pattern} of height $2n$---though technically speaking having $2n+1$ rows. Since we will make no use of odd orthogonal tableaux in the sequel, we leave the details to the interested reader.

Given an integer partition $\lambda$ with $\ell(\lambda) \leq n$, we denote by $SpGT_{2n}(\lambda)$ (respectively $OOGT_{2n}(\lambda)$) the set of all symplectic (respectively odd orthogonal) Gelfand--Tsetlin patterns of height $2n$ and shape $\lambda$. 

Let $h_k(x_1, x_2, \dots)$ be the $k$-th complete symmetric function defined (among many possibilities) by its generating series $\sum_{k \geq 0} h_k(x_1, x_2, \dots) z^k = \prod_{i} \frac{1}{1-x_i z}$. The \emph{Schur polynomials}~\cite{mac} and \emph{symplectic and odd orthogonal characters}~\cite{fh, sun2} can be defined by the following Jacobi--Trudi formulae:
\begin{equation} \label{eq:det_def}
    \begin{split}
    s_{\lambda} (x_1, \dots, x_n) &= \det [h_{\lambda_i - i + j}(x_1, \dots, x_n)]_{1 \leq i, j \leq \ell(\lambda)}, \\
    sp_{\lambda} (x_1^\pm, \dots, x_n^\pm) &= \frac{1}{2} \det [h_{\lambda_i - i + j}(x_1, x_1^{-1}, \dots, x_n, x_n^{-1}) + h_{\lambda_i - i - j + 2}(x_1, x_1^{-1}, \dots, x_n, x_n^{-1})]_{1 \leq i, j \leq \ell(\lambda)}, \\
    so^{\mathrm{odd}}_{\lambda}(x_1^\pm, \dots, x_n^\pm) &= \det [h_{\lambda_i - i + j}(x_1, x_1^{-1}, \dots, x_n, x_n^{-1}, 1) - h_{\lambda_i - i - j}(x_1, x_1^{-1}, \dots, x_n, x_n^{-1}, 1)]_{1 \leq i, j \leq \ell(\lambda)}.
    \end{split}
\end{equation}
The Schur polynomial is a manifestly symmetric polynomial in its variables, while the symplectic and orthogonal characters are Laurent polynomials having $BC$-symmetry---symmetry under permuting as well as \emph{inverting} the variables---and this explains the notation $x_i^{\pm}$. An equivalent way to define them---see~\cite{fk} for a combinatorial proof, and the way they will appear in this note, is as generating series of semi-standard (SSYTs), symplectic (SpTs) and odd orthogonal (OOTs) tableaux (Gelfand--Tsetlin patterns):
\begin{equation} \label{eq:tab_def}
    \begin{split}
    s_{\lambda} (x_1, \dots, x_n) &= \sum_{\substack{T:\text{\ SSYT} \\ \text{of shape\ } \lambda}} \prod_{i=1}^n x_i^{\text{number\ of\ $i$\ in\ $T$}} = \sum_{z \in GT_n(\lambda)} \prod_{i=1}^n x_i^{\mathrm{type}(z)_i}, \\
    sp_{\lambda} (x_1^\pm, \dots, x_n^\pm) &= \sum_{\substack{T:\text{\ SpT} \\ \text{of shape\ } \lambda}} \prod_{i=1}^n x_i^{\text{number\ of\ $i$\ in\ $T$} - \text{number\ of\ $\overline{i}$\ in\ $T$}} = \sum_{z \in SpGT_{2n}(\lambda)} \prod_{i=1}^n x_i^{\mathrm{type}(z)_{2i-1} - \mathrm{type}(z)_{2i}}, \\
    so^{\mathrm{odd}}_{\lambda} (x_1^\pm, \dots, x_n^\pm) &= \sum_{\substack{T:\text{\ OOT} \\ \text{of shape\ } \lambda}} \prod_{i=1}^n x_i^{\text{number\ of\ $i$\ in\ $T$} - \text{number\ of\ $\overline{i}$\ in\ $T$}} = \sum_{z \in OOGT_{2n}(\lambda)} \prod_{i=1}^n x_i^{\mathrm{type}(z)_{2i-1} - \mathrm{type}(z)_{2i}}.
    \end{split}
\end{equation}
For a combinatorial proof of the above equivalence, see~\cite{fk}.

\subsection{Proof of the main result}
\label{sec:main_proof}

We first prove Theorem~\ref{thm:main_thm}. The proof consists of four steps, outlined below.

\paragraph{Step 1.} The first step consists in showing that
\begin{equation} \label{eq:step1}
    \sum_{W \in \mathcal{W}^{\mathrm{p2hlr}}_{n, u}} \mathrm{wt}(W) = \left( \prod_{i=1}^n x_i \right)^u \sum_{\lambda: \lambda_1 \leq u} sp_{\lambda} (x_1^\pm, \dots, x_n^\pm). 
\end{equation}
This step is a discrete adaptation of the (finite temperature) argument given by Bisi and Zygouras~\cite{bz1}, and the proof also appears in Bisi's thesis~\cite[Theorem 3.5]{bis}. In can be summarized as follows: on a triangle $W \in \mathcal{W}^{\mathrm{p2hlr}}_{n, u}$ we inductively apply the bijective $rowRSK$ local rule of Appendix~\ref{sec:appendix_rsk} to obtain a \emph{generalized (semi-standard) oscillating tableau}---the entries of which will all be $\leq u$---and we subtract the entries of this tableau from $u$ to obtain a triangle of integers (between $0$ and $u$) which turns out to be our sought symplectic Gelfand--Tsetlin pattern. Its shape $\lambda$ has $\lambda_1 \leq u$. The transformation sends the weight $\mathrm{wt}(W)$ to the proper weight appearing in the tableau definition of the desired symplectic character $sp_{\lambda} (x_1^\pm, \dots, x_n^\pm)$. We note that the first part of the bijection is well-known and appears in e.g. the work of Krattenthaler~\cite{kra2}.

\begin{figure}[!ht]
    \begin{center}
    \includegraphics[scale=0.8]{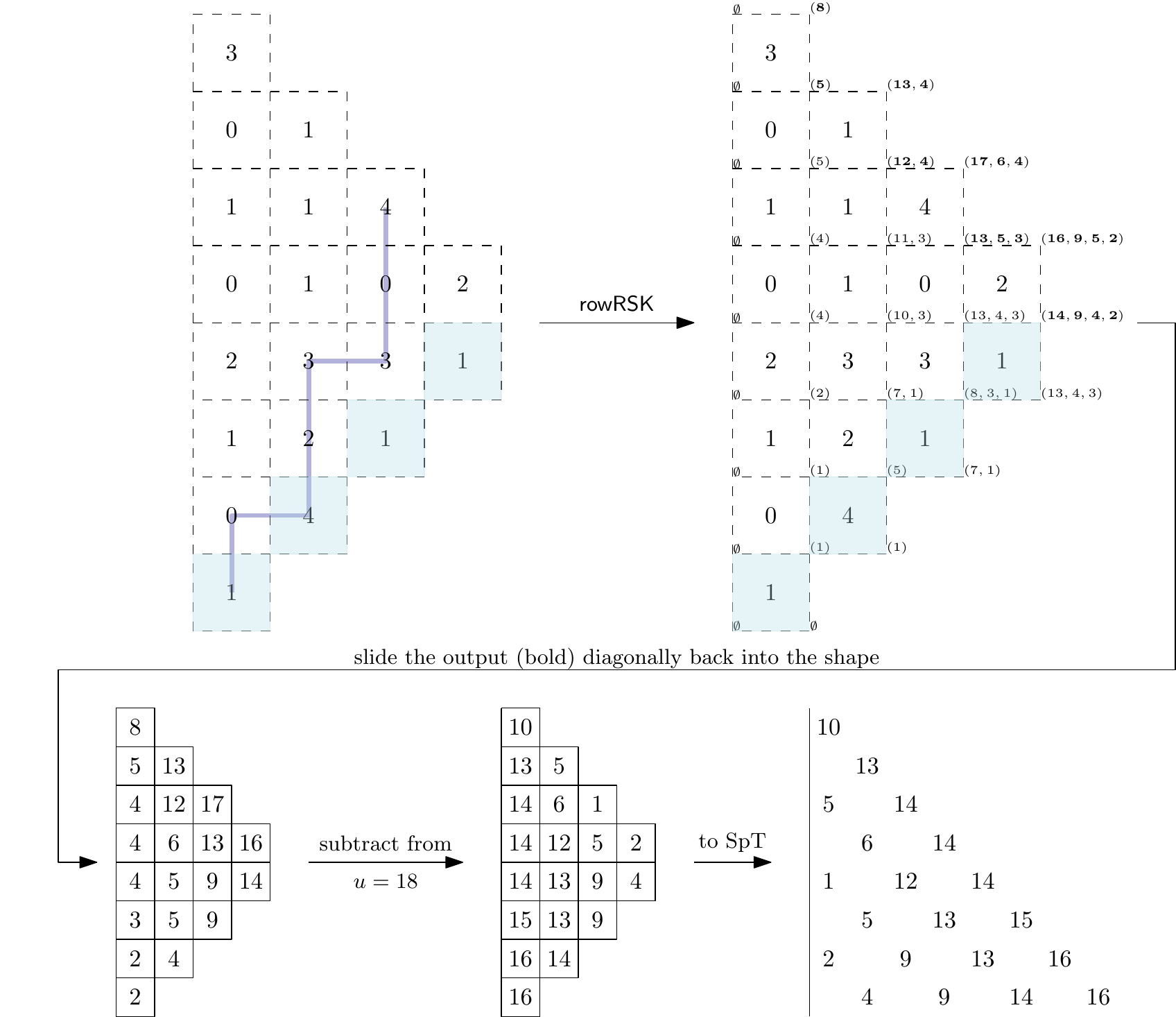}
    \end{center} 
    \caption{An example for the Bisi--Zygouras bijective procedure proving equation~\eqref{eq:step1} of Section~\ref{sec:main_proof}, going from an input triangle $W \in \mathcal{W}^{\mathrm{p2hlr}}_{n, u}$ for $(n, u) = (4, 18)$ (top left, with longest polymer highlighted in dark blue) via $rowRSK$ Fomin growth diagrams (top right, with output in bold) to a generalized oscillating tableau (bottom left) and then via entry-wise subtraction from $u$ to a symplectic tableau (bottom, middle and right).}
    \label{fig:p2hlr}
\end{figure}

More precisely, we take $W$---its entries sitting on $(\mathbb{N}+1/2)^2$ as depicted in Figure~\ref{fig:p2hlr}---and to it we inductively attach partitions $\mu^{(i, j)}$ sitting on the lattice points of $\mathbb{N}^2 \cap D^{\mathrm{p2hlr}}_n$. We assign the empty partition $\emptyset$ on the axes: $\mu^{(0, j)} = \mu^{(i, 0)} = \emptyset, i \in \{0, 1 \}, 0 \leq j \leq 2n$. To each other lattice point $(i, j)$---starting with $(1,1)$---we inductively assign the partition $\mu^{(i,j)}$ which is the output $\nu$ (in the notation of Appendix~\ref{sec:appendix_rsk}) of the $rowRSK$ local rule from Appendix~\ref{sec:appendix_rsk} applied on input $\alpha = \mu^{(i-1, j)}, \beta = \mu^{(i, j-1)},  \kappa = \mu^{(i-1, j-1)}, G = w_{i, j}$. We call $\mathsf{rowRSK}$ the totality of steps applying the local rule $rowRSK$ on $W$. We note that for a square $(i, i)$ on the $i=j$ diagonal (blue squares in Figure~\ref{fig:p2hlr}), once we have obtained $\kappa = \mu^{(i-1, i)}$ and $\alpha = \mu^{(i-1, i)}$ we set $\mu^{(i, i-1)} = \mu^{(i-1, i)}$ to act as our $\beta$ for the local rule $rowRSK$. At the end, the output of $\mathsf{rowRSK}$ is a sequence of up-down interlacing partitions
\begin{equation} \label{eq:oscillating}
    \emptyset = \mu^{(0, 2n)} \prec \mu^{(1, 2n)} \succ \mu^{(1, 2n-1)} \prec \dots \prec \mu^{(n, n+1)} \succ \mu^{(n, n)}
\end{equation}
depicted in bold in Figure~\ref{fig:p2hlr} (top right, the north and east borders). We note by construction $\ell(\mu^{(i,j)}) \leq \min(i,j)$ for all meaningful $i, j$. We then slide the partitions (except for the two empty ones) from equation~\eqref{eq:oscillating} back into the triangular shape: each partition on the north-east border goes south-west into the diagonal squares from the diagonal it lies on---see Figure~\ref{fig:p2hlr} (top right to bottom left). This is the generalized oscillating tableau (fig. cit., bottom left) and we note, due to Greene's Theorem~\ref{thm:greene}(a), that all its entries are $\leq u$ so we can subtract all said entries from $u$ and obtain a triangular tableau of numbers which now has weakly decreasing rows (left to right) and weakly increasing columns (top to bottom)---see Figure~\ref{fig:p2hlr} (bottom, middle panel). This is our symplectic Gelfand--Tsetlin pattern $z$ of height $2n$: more precisely, the $k$-th row of $z$ is the $i = j - (2n - k)$-th diagonal of the tableau just described---bottom last two panels in Figure~\ref{fig:p2hlr}. Its shape $\lambda:= z_{2n, \cdot}$ is the partition sitting on the diagonal $i=j$. 

Note that everything described so far is reversible: $rowRSK$ and thus $\mathsf{rowRSK}$ is a bijection, as is subtraction from~$u$. We now argue in more detail that the weights are preserved as desired.

Let $T=(t_{i, j})$ be the image under $\mathsf{rowRSK}$ of our array $W$---the generalized oscillating tableau depicted in Figure~\ref{fig:p2hlr} (bottom left). In terms of the sequence of partitions from~\eqref{eq:oscillating} we have
\begin{equation} \label{eq:mu_t}
    \mu^{\left( \lceil k/2 \rceil, 2n - \lfloor k/2 \rfloor  \right)}_{\lceil k/2 \rceil - i + 1} = t_{i, 2n - k + i}, \qquad 1 \leq i \leq \lceil k/2 \rceil, 1 \leq k \leq 2n. 
\end{equation}
Let $\mathrm{row}_i$ (respectively $\mathrm{col}_i$) be the sum of the $i$-th row (respectively column) of $W$. Inductive (repeated) application of~\eqref{eq:lc2}---see Remark~\ref{rem:size}---implies
\begin{equation}
    \mathrm{row}_{2n-i+1} = |\mu^{(i, 2n-i+1)}| - |\mu^{(i, 2n-i)}|, \qquad \mathrm{col}_i + \mathrm{row}_i - w_{i, i} = |\mu^{(i, 2n-i+1)}| - |\mu^{(i-1, 2n-i+1)}|, \qquad 1 \leq i \leq 2n. 
\end{equation}
Furthermore Greene's Theorem~\ref{thm:greene}(a) ensures $0 \leq t_{i, j} \leq u$ for all meaningful $i, j$. Let us change variables by setting 
\begin{equation} \label{eq:t_z}
    z_{i, j} := u - t_{j, 2n-i+j}, \qquad 1 \leq j \leq \lceil i/2 \rceil, 1 \leq i \leq 2n. 
\end{equation}
These variables also satisfy $0 \leq z_{i,j} \leq u$ for all meaningful $i, j$. Observe that because of the up-down interlacing constraints satisfied by the diagonals of $T$ (the partitions in \eqref{eq:oscillating}) and the further entry-wise subtraction from $u$, $z$ satisfies the interlacing constraints of a symplectic Gelfand--Tsetlin pattern. Its shape $\lambda$ clearly satisfies $\lambda_1 \leq u$. 

Denote by $|z_i| := \sum_{j=1}^{ \lceil i/2 \rceil} z_{i,j}$, i.e. the sum of the $i$-th row of $z$. The definition of $T$ and~\eqref{eq:t_z} implies that for all $1 \leq i \leq n$ we have
\begin{equation}
    |\mu^{(i, 2n-i+1)}| - |\mu^{(i-1, 2n-i+1)}| = u - |z_{2i-1}| + |z_{2i-2}|, \qquad |\mu^{(i, 2n-i+1)}| - |\mu^{(i, 2n-i)}| = |z_{2i}| - |z_{2i-1}|.
\end{equation}
 
Putting it all together, we obtain that
\begin{equation}
    \begin{split}
 \sum_{W \in \mathcal{W}^{\mathrm{p2hlr}}_{n, u}} \mathrm{wt}(W) &= \sum_{W \in \mathcal{W}^{\mathrm{p2hlr}}_{n, u}} \prod_{i=1}^n x_i^{\mathrm{row}_{2n+1-i} + \mathrm{col}_i + \mathrm{row}_i - w_{i,i} } \\
 &= \sum_{\substack{\text{oscillating tableau } T\\ \text{of the form~\eqref{eq:oscillating}}}} \prod_{i=1}^n x_i^{2 |\mu^{(i, 2n-i+1)}| - |\mu^{(i, 2n-i)}| - |\mu^{(i-1, 2n-i+1)}|} \\ 
 &= \sum_{\lambda:\lambda_1 \leq u} \sum_{ z \in SpGT_{n}(\lambda)} \prod_{i=1}^n x_i^{u - |z_{2i-1}| + |z_{2i-2}| - |z_{2i-1}| + |z_{2i}|} \\
 &= \left( \prod_{i=1}^n x_i \right)^u \sum_{\lambda:\lambda_1 \leq u} \sum_{z \in SpGT_{n}(\lambda)} \prod_{i=1}^n x_i^{-[\mathrm{type}(z)_{2i-1} - \mathrm{type}(z)_{2i})]} \\
 &= \left( \prod_{i=1}^n x_i \right)^u \sum_{\lambda: \lambda_1 \leq u} sp_{\lambda} (x_1^{\pm}, \dots, x_n^{\pm})  
    \end{split}
\end{equation}
where we note in the last equation we get the symplectic characters in the ``inverse variables'' $1/x_i$. Recalling $sp$ is symmetric under inversion of variables, we conclude the argument.

\paragraph{Step 2.} The next step consists in observing the following formula due to Okada~\cite{oka2}---see also~\cite[Theorem 4.3]{bkw} for a recent proof, notation similar to ours, and a wider context and use for formulas of this type:
\begin{equation}\label{eq:step2}
    \sum_{\lambda: \lambda_1 \leq u} sp_{\lambda} (x_1^\pm, \dots, x_n^\pm) = sp_{v^n} (x_1^\pm, \dots, x_n^\pm) so^{\mathrm{odd}}_{v^n} (x_1^\pm, \dots, x_n^\pm).
\end{equation}

\begin{rem}
    Equation~\eqref{eq:step2} holds for $u$ odd as well and reads
    \begin{equation}
        \sum_{\lambda: \lambda_1 \leq u} sp_{\lambda} (x_1^\pm, \dots, x_n^\pm) = sp_{s^n} (x_1^\pm, \dots, x_n^\pm) so^{\mathrm{odd}}_{t^n} (x_1^\pm, \dots, x_n^\pm)
    \end{equation}
    with $(s, t) = (\lfloor u/2 \rfloor, \lceil u/2 \rceil)$ and indeed this implies our assumption that $u$ was even was merely cosmetic to avoid floors and ceilings.
\end{rem}

\paragraph{Step 3.} The third step consists of rewriting the symplectic and odd orthogonal characters on the right-hand side above as bounded \emph{Littlewood sums} of Schur polynomials, combinatorially proven by Stembridge~\cite[Corollary 7.4]{ste}:
\begin{equation} \label{eq:step3}
\begin{split}
    \sum_{\lambda: \lambda_1 \leq u} s_{\lambda}(x_1, \dots, x_n) &= \left( \prod_{i=1}^n x_i \right)^v so^{\mathrm{odd}}_{v^n} (x_1^\pm, \dots, x_n^\pm), \\
    \sum_{\substack{\mu: \mu_1 \leq u,\\ \mu \text{ has even rows}}} s_{\mu}(x_1, \dots, x_n) &= \left( \prod_{i=1}^n x_i \right)^v sp_{v^n} (x_1^\pm, \dots, x_n^\pm).
\end{split}
\end{equation}

\paragraph{Step 4.} This fourth and final step consists of simply observing the following well-known identities~\cite{br1, PF04, fr} equating the bounded Littlewood sums with the desired generating series of polymers:
\begin{equation} \label{eq:step4}
    \begin{split}
        \sum_{W \in \mathcal{W}^{\mathrm{p2pr}}_{n, u}} \mathrm{wt}(W) &= \sum_{\lambda: \lambda_1 \leq u} s_{\lambda}(x_1, \dots, x_n), \\
        \sum_{W \in \mathcal{W}^{\mathrm{p2l}}_{n, v}} \mathrm{wt}(W) &= \sum_{\substack{\mu: \mu_1 \leq u,\\ \mu \text{ has even rows}}} s_{\mu}(x_1, \dots, x_n).
    \end{split}
\end{equation}
The proofs of both are standard applications of the Robinson--Schensted--Knuth algorithms described in the appendix. We omit the proof of the first (see~\cite{br1, fr} for details) but nevertheless prove the second equation using the seldom-used $colRSK$ local growth rule from the appendix. While our proof is equivalent to those in~\cite{br1, PF04, fr}, we chose to present because it uses a \emph{different} local growth rule than in the aforementioned references. Furthermore, the proof shows that ``the other RSK algorithm'' ($colRSK$) should be treated on an equal footing with the classical one of Knuth~\cite{knu} ($rowRSK$ in our language) and using it can lead to interesting observables in the theory of last passage percolation---in our language, to the generating series of (bounded by $v$) point-to-line polymers.

\begin{figure}[!ht]
    \begin{center}
    \includegraphics[scale=0.8]{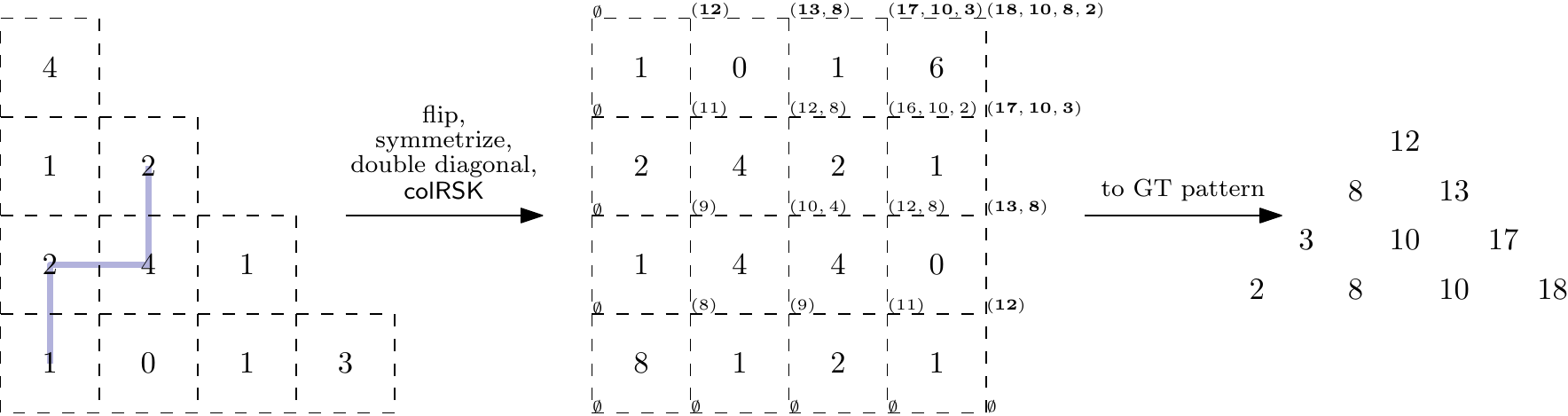}
    \end{center} 
    \caption{An example for the bijective procedure proving the second equation in~\eqref{eq:step1}, going from an input triangle $W \in \mathcal{W}^{\mathrm{p2l}}_{n, u}$ for $n=4, v \geq 9$ (left, with longest polymer of length $9$ highlighted in dark blue) via flipping/symmetrizing/doubling diagonal followed by the application of $colRSK$ Fomin local growth rules (middle, with output in bold) to a semi-standard Young tableau (right) of shape the even partition $(18, 10, 8, 2)$.} 
    \label{fig:p2l}
\end{figure}

First, take a triangle $W \in \mathcal{W}^{\mathrm{p2l}}_{n, v}$, flip it upside-down, double the numbers on the hypotenuse, and reflect the result across the hypotenuse to make it into a symmetric matrix we call $M = (M_{i,j})$ with indexing \emph{starting with the bottom left corner} which contains the $(1,1)$ entry---see first two panels of Figure~\ref{fig:p2l} for an example. Clearly this step is bijective.

Second, on $M$, perform the local rule $colRSK$ inductively similarly to what was done above. The matrix entries $M_{i, j}$ sit at half-integer lattice points $\left(i-1/2, j-1/2\right)$. On the integer points $0 \leq i, j \leq n$, place partitions $\mu^{(i, j)}$, starting the empty partition on the axes $i=0, 0 \leq j \leq n$ and $j=0, 0 \leq i \leq n$. For the other $1 \leq i, j \leq n$, proceed inductively---starting with $(i=1, j=1)$---as follows: place at $(i,j)$ the partition $\mu^{(i,j)}$ which is the output/result $\nu$ (in the notation of Appendix~\ref{sec:appendix_rsk}) of applying $colRSK$ on the input $\alpha = \mu^{(i-1, j)}, \beta = \mu^{(i, j-1)}, \kappa = \mu^{(i-1, j-1)}, G = M_{i, j}$. Let us call $\mathsf{colRSK}$ the inductive application of the $colRSK$ local rule $n^2$ times in the aforementioned way---see Figure~\ref{fig:p2l} (middle) for an example. It produces as output (depicted in bold in fig. cit.) the sequence of interlacing partitions on the outer north and east boundary of $M$:
\begin{equation} \label{eq:colRSK_output}
    \emptyset = \mu^{(0, n)} \prec \mu^{(1, n)} \prec \dots \prec \mu^{(n-1, n)} \prec \mu^{(n, n)} \succ \mu^{(n, n-1)} \succ \dots \succ \mu^{(n, 1)} \succ \mu^{(n, 0)} = \emptyset.
\end{equation}
As the matrix $M$ is symmetric and $colRSK$ from Appendix~\ref{sec:appendix_rsk} is manifestly symmetric in $\alpha$ and $\beta$, the sequence from~\eqref{eq:colRSK_output} is symmetric about its middle so the two Gelfand--Tsetlin patterns (equivalently SSYTs) corresponding to it are the same. Let us call the resulting pattern $z=(z_{i, j})_{1 \leq i \leq n, 1 \leq j \leq i}$: i.e., $z_{i, j} = \mu^{(n, j)}_i$. The shape of $z$, denoted $\mu$, is $\mu:= \mu^{(n, n)}$.

This step is also bijective as given the output partitions in equation~\eqref{eq:colRSK_output}, we can inductively apply $colRSK^{-1}$ from Appendix~\ref{sec:appendix_rsk} to obtain the matrix $M$ and the empty partitions on the axes (which of course we can then remove).  

Thus the output of our procedure, starting from $W$, is the GT pattern $z$: $W \longrightarrow z$. Due to Greene's Theorem~\ref{thm:greene}(b), all parts of $\mu$---the shape of $z$---are even as our matrix $M$ is symmetric and has even diagonal. Moreover, the same theorem yields $\mu_1 \leq 2v = u$ as $M$ comes from $W \in \mathcal{W}^{\mathrm{p2l}}_{n, v}$ on which we have, by definition, imposed exactly the conditions making this true. 

We finally argue that indeed the weights are preserved under the mapping $W \longrightarrow z$ when we pass to generating series. Let $\mathrm{col}_i^W$ (respectively $\mathrm{row}_i^W$) be the sum of the integers in the $i$-th column (respectively row) of $W$, and define $\mathrm{col}_i^M, \mathrm{row}_i^M$ similarly for $M$. By symmetry of $M$, $\mathrm{col}_i^M = \mathrm{row}_i^M$. Our definition of $M$ (recall the flipping that was involved) implies $2(\mathrm{col}_i^W + \mathrm{row}_{n-i+1}^W) = \mathrm{col}_i^M + \mathrm{row}_i^M = 2 \mathrm{col}_i^M$. Inductive application of~\eqref{eq:lc2}---see Remark~\ref{rem:size}---implies that for the symmetric interlacing sequence~\eqref{eq:colRSK_output}, we have
\begin{equation}
    \mathrm{col}_i^M = |\mu^{(i, n)}| - |\mu^{(i-1, n)}| = \mathrm{type}(z)_i
\end{equation}
Putting it all together using the notation $\mu: = \mu^{(n,n)} = \text{shape of } z$, we have:
\begin{equation}
    \begin{split}
    \sum_{W \in \mathcal{W}^{\mathrm{p2l}}_{n, v}} \mathrm{wt}(W) &= \sum_{W \in \mathcal{W}^{\mathrm{p2l}}_{n, v}} \prod_{i=1}^n x_i^{\mathrm{col}_i^W + \mathrm{row}_{n+1-i}^W}  \\
    &= \sum_{M} \prod_{i=1}^n x_i^{\mathrm{col}_i^M} \\
    &= \sum_{\substack{\mu:\mu_1 \leq u,\\ \mu \text{ has even rows}}} \sum_{z \in GT_n(\mu)} \prod_{i=1}^n x_i^{\mathrm{type}(z)_i} \\
    &= \sum_{\substack{\mu:\mu_1 \leq u,\\ \mu \text{ has even rows}}} s_{\mu} (x_1, \dots, x_n)
\end{split}
\end{equation}
and the result follows.

This concludes the proof of Theorem~\ref{thm:main_thm}. We now proceed to the proof of Corollary~\ref{cor:main_cor}. 

Suppose now that $0 < x_i < 1$ for all $i$ and that for each geometry the weights $w_{i, j}$ are geometric random variables of parameter $x_i x_j$ off-diagonal and $x_i$ on-diagonal---the later only for the p2hlr and p2pr geometries. We have just proven that:
\begin{equation}
    \label{eq:prob}
    \begin{split}
    \Prob(L^{\mathrm{p2hlr}}_n \leq u) &= \prod_{i=1}^n (1-x_i) \prod_{1 \leq i < j \leq n} (1-x_i x_j) \prod_{1 \leq i \leq j \leq n} (1-x_i x_j) \sum_{W \in \mathcal{W}^{\mathrm{p2hlr}}_{n, u}} \mathrm{wt}(W) \\
    &= \left( \prod_{i=1}^n (1-x_i) \prod_{1 \leq i < j \leq n} (1-x_i x_j) \sum_{W \in \mathcal{W}^{\mathrm{p2pr}}_{n, v}} \mathrm{wt}(W) \right)  \left( \prod_{1 \leq i \leq j \leq n} (1-x_i x_j) \sum_{W \in \mathcal{W}^{\mathrm{p2l}}_{n, v}} \mathrm{wt}(W) \right) \\
    &= \Prob(L^{\mathrm{p2pr}}_n \leq u) \cdot \Prob(L^{\mathrm{p2l}}_n \leq v).
    \end{split}
\end{equation}
Fix $0 < q < 1$ and put all $x_i = \sqrt{q}$. Then if above on the right-hand side we take $u = c_1 n + c_2 n^{1/3} s$ with $c_{1}$ and $c_2$ from ~\eqref{eq:constants}, both probabilities (recall $v = u/2$) have the same well-defined $n \to \infty$ limit---see~\cite{br99, br2, si, PF04, BBCS17, bz2, bbnv} for various proofs but note some are of the results only in the limit $q \to 0$ or $q \to 1$. The common limit is the Tracy--Widom GOE~\cite{TW96} distribution function $F_1(s)$. The result follows.

\section{Conclusion}
\label{sec:conclusion}

We conclude with a few remarks. First, the proof of Theorem~\ref{thm:main_thm} is not completely and transparently bijective. The equations in steps 1 and 4~\eqref{eq:step1},~\eqref{eq:step4} are consequences of RSK variants, and even the second equation in step 3~\eqref{eq:step3} can be seen as such\footnote{We address this and its connection to last passage percolation and other combinatorial models in upcoming work in progress with Bisi and Zygouras.}. We think the first identity of~\eqref{eq:step3} is also a consequence of (some) RSK but know of no proof. Nonetheless, we suspect proving~\eqref{eq:step2} using an RSK variant is significantly harder, as we would be proving a \emph{Littlewood--Richardson}-type rule (albeit one for very simple shapes). Dually, it would be interesting to have a probabilistic interpretation/derivation of~\eqref{eq:prob}---the probabilistic version of Theorem~\ref{thm:main_thm}---even in the iid case $x_i = \sqrt{q} \ \forall i$.

Second, while formulas of type~\eqref{eq:step1} and~\eqref{eq:step2} can be characterized as ``representation-theoretic happy accidents'', the question that arises is if there exist other such accidents that are useful for the analysis of different LPP variants (or other mathematical physics) models. A small catalogue of useful such formulas can be found in the recent works of Brent--Krattenthaler--Warnaar~\cite{bkw} and Rains--Warnaar~\cite{rw}. 

Third, one can presumably use the techniques of Betea--Bouttier--Nejjar--Vuleti\'c~\cite{bbnv} to obtain not just one-point distributional results and asymptotics for the p2hlr polymer but also full multi-point correlations, as double contour integrals, for the whole process---i.e., distributions for collections of non-intersecting longest up-right paths. One can additionally add extra parameters to the model, like an $\alpha$ parameter governing the diagonal---see~\cite[Section 6]{fr}. It is known that $\alpha$ governs the transition between GUE and $\mathrm{GOE}^2$ fluctuations~\cite{br99, br2} at the one-point distributional level. Is there anything interesting happening at the multi-point level?

Fourth, it is tempting to speculate whether Theorem~\ref{thm:main_thm} has a finite temperature (geometric/Whittaker) analogue. Indeed it was in this context that the Whittaker equivalent of equation~\eqref{eq:step1} was discovered by Bisi and Zygouras~\cite{bz1}. The main obstruction we see is a finite temperature analogue of~\eqref{eq:step2}. However the existence of a Macdonald--Koornwinder lift to eq. cit. would go a long way as one could presumably take appropriate limits towards the desired goal. A finite temperature analogue of Corollary~\ref{cor:main_cor} seems further away as there are no known rigorous asymptotical results for the logarithm of the partition function of the O'Connell--Yor point-to-line gamma polymer.

Last, some of the equations used in Section~\ref{sec:main_proof} along with both RSK correspondences themselves have $t$ lifts (deformations) to the Hall--Littlewood level. For example, equation~\eqref{eq:step2} (at least for $v=\infty$) can be recast as a partition function identity in an appropriate 6-vertex model~\cite{wzj}; equation~\eqref{eq:step3} has a plethora of Hall--Littlewood (even Macdonald or elliptic) analogues~\cite{rw}; and the RSK correspondences from the appendix have $t$ lifts as well, recently discovered by Bufetov--Matveev~\cite{bm}. Can one piece together these facts to obtain distributional results for height functions in the appropriate 6-vertex model? Something similar has already been achieved in a simpler setting by Borodin--Bufetov--Wheeler~\cite{bbw}.

\appendix

\section{Fomin growth and Greene's theorem}
\label{sec:appendix_rsk}

We recall here, for the benefit of the reader and in the form of Fomin growth diagrams, the two Robinson--Schensted--Knuth bijections used throughout the note. 

Fix two partitions $\alpha, \beta$. We list two $(\max, +)$ local Fomin growth rules, dubbed $rowRSK$ and $colRSK$, which provide bijections between the two sets 
\begin{equation}\label{eq:lc1}
    \{\kappa \text{ a partition}: \alpha \succ \kappa \prec \beta \} \times \{G: G \in \mathbb{N} \} \longleftrightarrow \{\nu \text{ a partition}: \alpha \prec \nu \succ \beta \}
\end{equation}
satisfying the condition
\begin{equation}
\label{eq:lc2}
    |\kappa| + |\nu| = |\alpha| + |\beta| + G.
\end{equation}
The rule $rowRSK$ is Fomin's~\cite{fom2, fom} local rule for the Robinson--Schensted--Knuth (RSK) \emph{row insertion} correspondence~\cite{knu}, as first written down by Gessel~\cite{ges2}. The second, $colRSK$, is a local growth rule for Burge's \emph{column insertion} RSK---see Appendix A.4 of~\cite{ful} and~\cite[page 21, second paragraph, after ``It can also be used...'']{bur}. It is, up to conjugating the partitions and other ``minor'' manipulations, the same rule as Krattenthaler's third variation from~\cite{kra} and equivalent to the Hilman--Grassl bijection---see~\cite{hg} but also the recent~\cite{ppm} for an extended explanation. For a dynamical view of both rules in terms of particles hopping on a lattice, see~\cite{bp, pm}---in particular we arrived at the $colRSK$ construction below by reverse-engineering the ``column $\alpha$'' algorithm therein. The rules are listed below, taking as inputs pairs $(\kappa, G)$ with $\kappa$ a partition and $G\in \mathbb{N}$---and of course the fixed partitions $\alpha, \beta$---and producing the partition~$\nu$.
\newline

\begin{tabular}{|l|l|}
    \hline
\begin{tabular}{@{}l@{}}
\noindent \textbf{Algorithm} $rowRSK$\\ \ \\
\textbf{Input:} $\alpha, \beta; \kappa, G$ satisfying $\alpha \succ \kappa \prec \beta$\\
$\ell - 1 = \min(\ell(\alpha), \ell(\beta))$\\
$\nu_1 = \max(\alpha_1, \beta_1) + G$\\
\textbf{FOR} $s = 2, 3, \dots, \ell$\\
\indent  $\nu_s = \max(\alpha_s, \beta_s) + \min(\alpha_{s-1}, \beta_{s-1}) - \kappa_{s-1}$\\
\textbf{ENDFOR}\\
\textbf{Output:} $\nu$ satisfying $\alpha \prec \nu \succ \beta$\\
\ \\
\
\end{tabular}
& 
\begin{tabular}{@{}l@{}}
\noindent \textbf{Algorithm} $colRSK$\\ \ \\
\textbf{Input:} $\alpha, \beta; \kappa, G$ satisfying $\alpha \succ \kappa \prec \beta$ \\
$\ell - 1 = \min(\ell(\alpha), \ell(\beta))$ \\
$G_{\ell} = G$ \\
\textbf{FOR} $s = \ell, \ell - 1, \dots, 1$  \\
\indent  $\nu_s = \min(\max(\alpha_s, \beta_s) + G_s, \kappa_{s-1})$ \\
\indent  $G_{s-1} = G_s - \min(G_s, \kappa_{s-1} - \max(\alpha_s, \beta_s))$ \\
\indent \indent \indent \indent $+ \min(\alpha_{s-1}, \beta_{s-1}) - \kappa_{s-1}$ \\
\textbf{ENDFOR} \\
\textbf{Output:} $\nu$ satisfying $\alpha \prec \nu \succ \beta$
\end{tabular}\\
    \hline
\end{tabular}
\newline

That $rowRSK$ and $colRSK$ produce $\nu$ satisfying $|\kappa| + |\nu| = |\alpha| + |\beta| + G$ is immediate from their explicit form, as are the interlacing conditions satisfied by $\nu$ given the ones satisfied by $\kappa$. Moreover, $rowRSK$ can be manifestly inverted thus showing it is indeed a bijection. For $colRSK$ and the benefit of the reader, we list the inverse below, taking $\nu$ and producing $(\kappa, G)$.
\newline

\begin{tabular}{|l|}
    \hline
\begin{tabular}{@{}l@{}}
\noindent \textbf{Algorithm} $colRSK^{-1}$\\ \ \\
\textbf{Input:} $\alpha, \beta; \nu$ satisfying $\alpha \prec \nu \succ \beta$\\
$\ell - 1 = \min(\ell(\alpha), \ell(\beta))$ \\
$G_{1} = \max(\alpha_1, \beta_1) - \nu_1$ \\
\textbf{FOR} $s = 1,2,\dots, \ell - 1$  \\
\indent  $\kappa_s = \max(\min(\alpha_s, \beta_s) - G_s, \nu_{s+1})$ \\
\indent  $G_{s+1} = \nu_{s+1} + G_s - \max(\alpha_{s+1}, \beta_{s+1}) - \min(\alpha_{s}, \beta_{s}) + \kappa_{s}$ \\
\textbf{ENDFOR} \\
$G = G_{\ell}$ \\
\textbf{Output:} $G, \kappa$ satisfying $\alpha \succ \kappa \prec \beta$ 
\end{tabular}\\
\hline
\end{tabular}
\newline

\begin{rem} \label{rem:skew_cauchy}
The interlacing conditions on $\kappa$ and $\nu$ preserved by both $rowRSK$ and $colRSK$ bijections~\eqref{eq:lc1} along with the size condition~\eqref{eq:lc2} imply that both bijections give proofs of the skew Cauchy identity
\begin{equation}
    \sum_{\nu} s_{\nu / \alpha} (x) s_{\nu / \beta} (y) = \frac{1}{1-x y}  \sum_{\kappa} s_{\alpha / \kappa} (y) s_{\beta / \kappa} (x)
\end{equation}
where $x, y$ are variables, and $s_{\lambda / \mu} (x) = x^{|\lambda| - |\mu|} \delta_{\mu \prec \lambda}$ is the skew Schur function specialized in a single variable. The same identity (in any number of variables) is also proven bijectively by Sagan and Stanley~\cite{ss2} using the skew tableau insertion version of $rowRSK$. The interested reader can easily modify their argument for a skew tableau insertion version of $colRSK$.
\end{rem}

Among the many bijections (local growth rules) that can be produced satisfying~\eqref{eq:lc1} and~\eqref{eq:lc2}, these two seem (to the best of our knowledge) to be the only ones satisfying a Greene-type~\cite{gre} theorem (see also~\cite{kra} for the $colRSK$ version). 

\begin{figure}[!ht]
    \centering
    \includegraphics[scale=0.7]{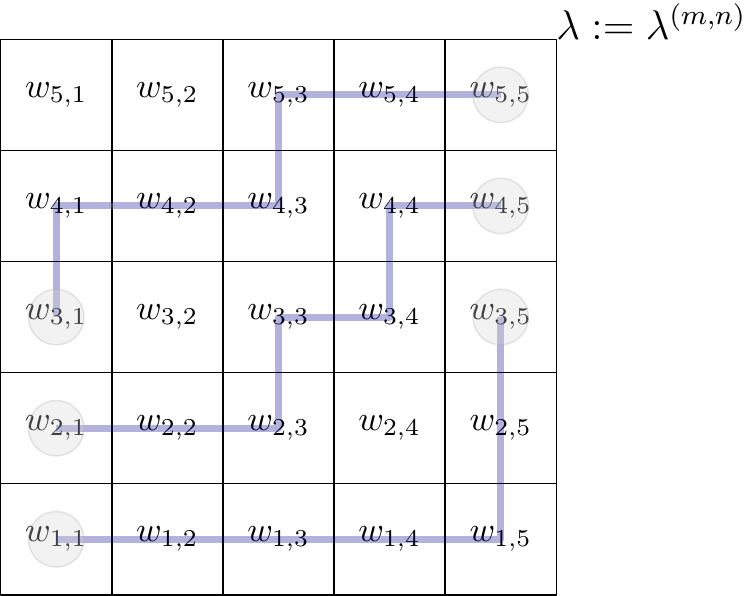} \qquad \qquad \includegraphics[scale=0.7]{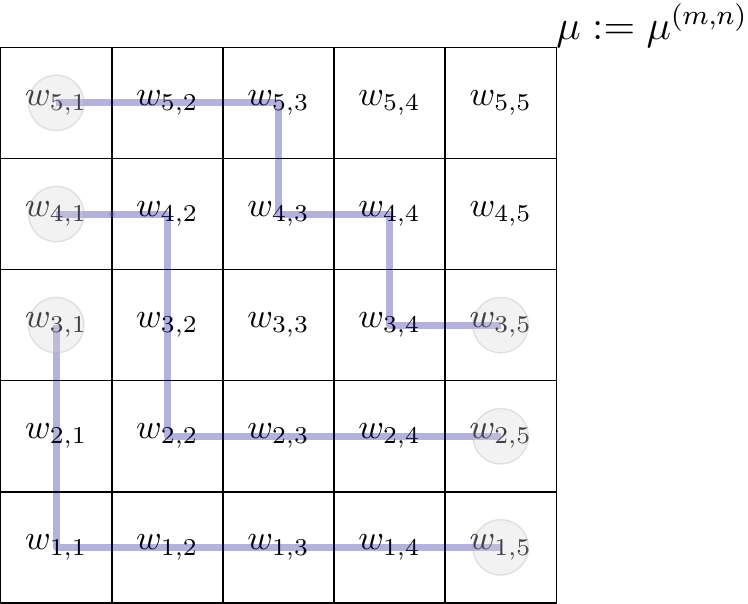}
    \caption{Examples of $k$ non-intersecting lattice paths appearing in the statement of Greene's Theorem~\ref{thm:greene} for $m=n=5, k=3$. Left: paths appearing in Theorem~\ref{thm:greene}(a) for $\mathsf{rowRSK}$. Right: paths appearing in Theorem~\ref{thm:greene}(b) for $\mathsf{colRSK}$. All squares are filled with non-negative integers $w_{i,j}$. The weight of any such collection of paths is the sum over all the numbers crossed by the paths. Starting/ending points are denoted by small disks.}
    \label{fig:greene_paths}
\end{figure}

To set up the picture, suppose we have non-negative integers $(w_{i,j})_{1 \leq i,j}$ sitting at the half-integer points $\left( i-1/2, j-1/2 \right)$ of $(\mathbb{N}+1/2)^2$ and consider the $m \times n$ matrix $(w_{i,j})_{1 \leq i \leq m, 1 \leq j \leq n}$ sitting inside the rectangle $R:=[0, m] \times [0, n]$---see Figure~\ref{fig:greene_paths}. On the integer points inside $R$ we put the empty partition on the axes and then put partitions $\lambda^{(i,j)}$ (respectively $\mu^{(i,j)}$) on the other lattice points---starting with $(1,1)$---which we define as the output $\nu$ of inductive successive applications of the rule $rowRSK$ (respectively $colRSK$) on input $\alpha = \lambda^{(i-1,j)}, \beta = \lambda^{(i,j-1)}, \kappa = \lambda^{(i-1,j-1)}, G = w_{i,j}$ (respectively $\alpha = \mu^{(i-1,j)}, \beta = \mu^{(i,j-1)}, \kappa = \mu^{(i-1,j-1)}, G = w_{i,j}$ for the case of $colRSK$). By construction $\ell(\lambda^{(i, j)}), \ell(\mu^{(i, j)}) \leq \min(i,j)$. Let us call $\mathsf{rowRSK}$ this sequence of $mn$ successive applications of $rowRSK$, and similarly for $\mathsf{colRSK}$ and $colRSK$. Let $\lambda:= \lambda^{(m, n)}$ (respectively $\mu:= \mu^{(m, n)}$) be the partition sitting at the outermost corner---see Figure~\ref{fig:greene_paths}---after applying $\mathsf{rowRSK}$ (respectively $\mathsf{colRSK}$). Pick a non-negative integer $k \leq \min (m, n)$. The following was proven by Greene~\cite{gre} and connects the RSK bijections herein described with last passage percolation models. 

\begin{thm}[Greene~\cite{gre}]\label{thm:greene}
(a) After application of $\mathsf{rowRSK}$ on input $(w_{i,j})_{1 \leq i \leq m, 1 \leq j \leq n}$, we have:
\begin{equation}
\lambda_1 + \dots + \lambda_k = \max_{\pi \in P_k} \sum_{\left(i, j \right) \in \pi} w_{i,j}
\end{equation} 
where $P_k$ is the collection of $k$ non--intersecting up-right paths (i.e., having unit steps up or east from one square to its adjacent), starting from the south--west vertical edge of the matrix and ending on the north--east vertical edge. In particular $\lambda_1$ is length of the the longest up-right path going from the south-west corner to the north-east corner of the rectangle $R$.

(b) Moreover, after application of $\mathsf{colRSK}$ on input $(w_{i,j})_{1 \leq i \leq m, 1 \leq j \leq n}$, we have:
\begin{equation}
\mu_1 + \dots + \mu_k = \max_{\pi \in P'_k} \sum_{\left(i, j \right) \in \pi} w_{i,j}
\end{equation} 
where $P'_k$ is the collection of $k$ non--intersecting down-right paths (i.e., having unit steps down or east from one square to its adjacent), starting from the north--west vertical edge of the matrix and ending on the south--east vertical edge. In particular $\mu_1$ is  length of the the longest down-right path going from the north-west to the south-east corners of $R$.
\end{thm}

Examples of $k$ non-intersecting paths from $P_k, P'_k$ appear in Figure~\ref{fig:greene_paths}. The proof that $\lambda_1$ is as stated is immediate from the $rowRSK$ local rule construction. We do not know of a direct proof for $\mu_1$ and $colRSK$, though one can of course reinterpret $colRSK$ in terms of the Burge column insertion~\cite{bur} and then proceed as was done by Greene~\cite{gre} using the so-called Knuth relations~\cite{knu}. 

\begin{rem}
    \label{rem:size}
    With the same setup as above, if $\mathrm{row}_k$ (respectively $\mathrm{col}_k$) denote the sum of the $k$-th row (respectively column) of $w$, inductive application of~\eqref{eq:lc2} yields 
    \begin{equation}
        \mathrm{row}_k = |\lambda^{(m, k)}| - |\lambda^{(m, k-1)}| = |\mu^{(m, k)}| - |\mu^{(m, k-1)}|, \qquad \mathrm{col}_k = |\lambda^{(k, n)}| - |\lambda^{(k-1, n)}| = |\mu^{(k, n)}| - |\mu^{(k-1, n)}|.
    \end{equation}
\end{rem}

\bibliographystyle{myhalpha}
\bibliography{GOE_sq}

\providecommand{\noopsort}[1]{}
\begin{thebibliography}{BBNV18}

\bibitem[BBCS17]{BBCS17}
J.~Baik, G.~Barraquand, I.~Corwin, and T.~Suidan.
\newblock Pfaffian {S}chur processes and last passage percolation in a
  half-quadrant.
\newblock {\em Ann. Probab.}, to appear, 2017, arXiv:1606.00525v3 [math.PR].

\bibitem[BBNV18]{bbnv}
D.~Betea, J.~Bouttier, P.~Nejjar, and M.~Vuleti\'c.
\newblock The free boundary {S}chur process and applications {I}.
\newblock {\em Ann. Henri Poincar\'e}, to appear, 2018, arXiv:1704.05809v2
  [math.PR].

\bibitem[BBW16]{bbw}
A.~Bufetov, A.~Borodin, and M.~Wheeler.
\newblock Between the stochastic six vertex model and {H}all--{L}ittlewood
  processes.
\newblock 2016, arXiv:1611.09486 [math.PR].

\bibitem[Bis18a]{bis}
E.~Bisi.
\newblock Random polymers via orthogonal {W}hittaker and symplectic {S}chur
  functions, {P}h{D} thesis.
\newblock {\em Department of Statistics, University of Warwick}, 2018,
  arXiv:1810.03734 [math.PR].

\bibitem[Bis18b]{bis2}
E.~Bisi.
\newblock Work in progress.
\newblock 2018.

\bibitem[BKW16]{bkw}
R.~P. Brent, C.~Krattenthaler, and S.~O. Warnaar.
\newblock Discrete analogues of {M}acdonald-{M}ehta integrals.
\newblock {\em J. Combin. Theory Ser. A}, 144:80--138, 2016.

\bibitem[BM17]{bm}
A.~Bufetov and K.~Matveev.
\newblock Hall--{L}ittlewood {RSK} field.
\newblock 2017, arXiv:1705.07169 [math.PR].

\bibitem[BP16]{bp}
A.~Borodin and L.~Petrov.
\newblock Nearest neighbor {M}arkov dynamics on {M}acdonald processes.
\newblock {\em Advances in Mathematics}, 300:71 -- 155, 2016.
\newblock Special volume honoring Andrei Zelevinsky.

\bibitem[BR99]{br99}
J.~Baik and E.~M. Rains.
\newblock Symmetrized random permutations, 1999, arXiv:math/9910019 [math.CO].

\bibitem[BR01a]{br1}
J.~Baik and E.~M. Rains.
\newblock {A}lgebraic aspects of increasing subsequences.
\newblock {\em Duke Math. J.}, 109(1):1--65, 2001, arXiv:math/9905083
  [math.CO].

\bibitem[BR01b]{br2}
J.~Baik and E.~M. Rains.
\newblock {T}he asymptotics of monotone subsequences of involutions.
\newblock {\em Duke Math. J.}, 109(2):205--281, 2001, arXiv:math/9905084
  [math.CO].

\bibitem[Bur74]{bur}
W.~H. Burge.
\newblock Four correspondences between graphs and generalized {Y}oung tableaux.
\newblock {\em J. Combinatorial Theory Ser. A}, 17:12--30, 1974.

\bibitem[BZ17a]{bz2}
E.~Bisi and N.~Zygouras.
\newblock {GOE} and {A}iry{$_{2\to 1}$} marginal distribution via symplectic
  {S}chur functions.
\newblock 2017, arXiv:1711.05120v1 [math.PR].

\bibitem[BZ17b]{bz1}
E.~Bisi and N.~Zygouras.
\newblock Point-to-line polymers and orthogonal {W}hittaker functions.
\newblock {\em Trans. Amer. Math. Soc.}, to appear, 2017, arXiv:1703.07337v4
  [math.PR].

\bibitem[Fer04]{PF04}
P.~Ferrari.
\newblock {P}olynuclear growth on a flat substrate and edge scaling of {GOE}
  eigenvalues.
\newblock {\em Comm. Math. Phys.}, 252(1):77--109, 2004.

\bibitem[FH91]{fh}
W.~Fulton and J.~Harris.
\newblock {\em Representation theory}, volume 129 of {\em Graduate Texts in
  Mathematics}.
\newblock Springer-Verlag, New York, 1991.
\newblock A first course, Readings in Mathematics.

\bibitem[FK97]{fk}
M.~Fulmek and C.~Krattenthaler.
\newblock Lattice path proofs for determinantal formulas for symplectic and
  orthogonal characters.
\newblock {\em J. Combin. Theory Ser. A}, 77(1):3--50, 1997.

\bibitem[Fom86]{fom2}
S.~Fomin.
\newblock The generalized {R}obinson-{S}chensted-{K}nuth correspondence.
\newblock {\em Zap. Nauchn. Sem. Leningrad. Otdel. Mat. Inst. Steklov. (LOMI)},
  155(Differentsial'naya Geometriya, Gruppy Li i Mekh. VIII):156--175, 195,
  1986.

\bibitem[Fom95]{fom}
S.~Fomin.
\newblock Schur operators and {K}nuth correspondences.
\newblock {\em J. Combin. Theory Ser. A}, 72(2):277--292, 1995.

\bibitem[FR07]{fr}
P.~J. Forrester and E.~M. Rains.
\newblock Symmetrized models of last passage percolation and non-intersecting
  lattice paths.
\newblock {\em J. Stat. Phys.}, 129(5-6):833--855, 2007.

\bibitem[Ful97]{ful}
W.~Fulton.
\newblock {\em Young tableaux}, volume~35 of {\em London Mathematical Society
  Student Texts}.
\newblock Cambridge University Press, Cambridge, 1997.
\newblock With applications to representation theory and geometry.

\bibitem[Ges93]{ges2}
I.~M. Gessel.
\newblock Counting paths in {Y}oung's lattice.
\newblock {\em J. Statist. Plann. Inference}, 34(1):125--134, 1993.

\bibitem[Gre74]{gre}
C.~Greene.
\newblock An extension of {S}chensted's theorem.
\newblock {\em Advances in Math.}, 14:254--265, 1974.

\bibitem[HG76]{hg}
A.~Hillman and R.~Grassl.
\newblock Reverse plane partitions and tableau hook numbers.
\newblock {\em Journal of Combinatorial Theory, Series A}, 21(2):216 -- 221,
  1976.

\bibitem[Joh00]{joh2}
K.~Johansson.
\newblock Shape fluctuations and random matrices.
\newblock {\em Comm. Math. Phys.}, 209(2):437--476, 2000, arXiv:math/9903134
  [math.CO].

\bibitem[KES83]{kel}
R.~C. King and N.~G.~I. El-Sharkaway.
\newblock Standard young tableaux and weight multiplicities of the classical
  lie groups.
\newblock {\em Journal of Physics A: Mathematical and General}, 16(14):3153,
  1983.

\bibitem[Kin76]{kin}
R.~C. King.
\newblock Weight multiplicities for the classical groups.
\newblock In A.~Janner, T.~Janssen, and M.~Boon, editors, {\em Group
  Theoretical Methods in Physics}, pages 490--499, Berlin, Heidelberg, 1976.
  Springer Berlin Heidelberg.

\bibitem[Knu70]{knu}
D.~E. Knuth.
\newblock {P}ermutations, matrices, and generalized {Y}oung tableaux.
\newblock {\em Pacific J. Math.}, 34:709--727, 1970.

\bibitem[Kra06]{kra}
C.~Krattenthaler.
\newblock {G}rowth diagrams, and increasing and decreasing chains in fillings
  of {F}errers shapes.
\newblock {\em Adv. in Appl. Math.}, 37(3):404--431, 2006.

\bibitem[Kra16]{kra2}
C.~Krattenthaler.
\newblock Bijections between oscillating tableaux and (semi)standard tableaux
  via growth diagrams.
\newblock {\em J. Combin. Theory Ser. A}, 144:277--291, 2016.

\bibitem[Mac95]{mac}
I.~G. Macdonald.
\newblock {\em {S}ymmetric functions and {H}all polynomials}.
\newblock Oxford Mathematical Monographs. The Clarendon Press Oxford University
  Press, New York, second edition, 1995.
\newblock With contributions by A. Zelevinsky, Oxford Science Publications.

\bibitem[MP17]{pm}
K.~Matveev and L.~Petrov.
\newblock {$q$}-randomized {R}obinson-{S}chensted-{K}nuth correspondences and
  random polymers.
\newblock {\em Ann. Inst. Henri Poincar\'e D}, 4(1):1--123, 2017.

\bibitem[MPP18]{ppm}
A.~H. Morales, I.~Pak, and G.~Panova.
\newblock Hook formulas for skew shapes {I}. q-analogues and bijections.
\newblock {\em Journal of Combinatorial Theory, Series A}, 154:350 -- 405,
  2018.

\bibitem[Oka98]{oka2}
S.~Okada.
\newblock Applications of minor summation formulas to rectangular-shaped
  representations of classical groups.
\newblock {\em J. Algebra}, 205(2):337--367, 1998.

\bibitem[Oko01]{oko}
A.~Okounkov.
\newblock {I}nfinite wedge and random partitions.
\newblock {\em Selecta Math. (N.S.)}, 7(1):57--81, 2001, arXiv:math/9907127
  [math.RT].

\bibitem[Rai00]{rai}
E.~M. Rains.
\newblock {C}orrelation functions for symmetrized increasing subsequences,
  2000, arXiv:math/0006097 [math.CO].

\bibitem[RW18]{rw}
E.~M. Rains and O.~Warnaar.
\newblock Bounded {L}ittlewood identities.
\newblock {\em Memoirs of the Amer. Math. Soc.}, to appear, 2018,
  arXiv:1506.02755v3 [math.CO].

\bibitem[SI04]{si}
T.~Sasamoto and T.~Imamura.
\newblock Fluctuations of the one-dimensional polynuclear growth model in
  half-space.
\newblock {\em J. Statist. Phys.}, 115(3-4):749--803, 2004,
  arXiv:cond-mat/0307011 [cond-mat.stat-mech].

\bibitem[SS90]{ss2}
B.~E. Sagan and R.~P. Stanley.
\newblock Robinson-{S}chensted algorithms for skew tableaux.
\newblock {\em J. Combin. Theory Ser. A}, 55(2):161--193, 1990.

\bibitem[Ste90]{ste}
J.~R. Stembridge.
\newblock Nonintersecting paths, {P}faffians, and plane partitions.
\newblock {\em Adv. Math.}, 83(1):96--131, 1990.

\bibitem[Sun90a]{sun2}
S.~Sundaram.
\newblock Tableaux in the representation theory of the classical {L}ie groups.
\newblock In {\em Invariant theory and tableaux ({M}inneapolis, {MN}, 1988)},
  volume~19 of {\em IMA Vol. Math. Appl.}, pages 191--225. Springer, New York,
  1990.

\bibitem[Sun90b]{sun3}
S.~Sundaram.
\newblock Orthogonal tableaux and an insertion algorithm for so(2n + 1).
\newblock {\em Journal of Combinatorial Theory, Series A}, 53(2):239 -- 256,
  1990.

\bibitem[TW94]{tw}
C.~A. Tracy and H.~Widom.
\newblock Level-spacing distributions and the {A}iry kernel.
\newblock {\em Comm. Math. Phys.}, 159(1):151--174, 1994, arXiv:hep-th/9211141.

\bibitem[TW96]{TW96}
C.~A. Tracy and H.~Widom.
\newblock On orthogonal and symplectic matrix ensembles.
\newblock {\em Communications in Mathematical Physics}, 177(3):727--754, Apr
  1996.

\bibitem[WZJ16]{wzj}
M.~Wheeler and P.~Zinn-Justin.
\newblock Refined {C}auchy/{L}ittlewood identities and six-vertex model
  partition functions: Iii. {D}eformed bosons.
\newblock {\em Advances in Mathematics}, 299:543 -- 600, 2016.

\end{thebibliography}

\pagebreak

\

\end{document}